\documentclass[12pt]{article}

\title{Buonomano against Bell: \\ Nonergodicity or nonlocality?}
\author{Andrei Khrennikov\\
International Center for Mathematical
Modeling \\
Linnaeus University, V\"axj\"o, S-35195, Sweden\\
National Research University of  Information Technologies,  \\Mechanics and Optics (ITMO),
St. Petersburg, Russia}

\begin{document}
\maketitle

\abstract{The aim of this note is to attract attention of the quantum foundational community to the fact  that in Bell's arguments one can not
distinguish two hypotheses:  a) quantum mechanics is {\it nonlocal}, b)  quantum mechanics is {\it  nonergodic.} Therefore
experimental violations of Bell's inequality can be as well interpreted as supporting the hypothesis  that  stochastic processes
induced by quantum measurements are  nonergodic.  The latter hypothesis was discussed actively by Buonomano since 1980th. However, in contrast 
to Bell's hypothesis on nonlocality it did not attract so much attention. The only experiment testing the hypothesis on nonergodicity was performed in neutron 
interferometry (by Summhammer, in 1989).  This experiment can be considered as rejecting  this hypothesis. However, it cannot be considered as a decisive experiment.
New experiments are badly needed. We point out that a nonergodic model can be realistic, i.e., the distribution of hidden (local!) variables is well defined.
We also discuss coupling of violation of the Bell inequality with violation of the condition of weak mixing for ergodic dynamical systems.
}

{\bf keywords:} Hypothesis of ergodicity, Bell inequality, time and ensemble averages, locality, realism, ontic and epistemic descriptions, ergodic theorems, weak mixing

\section{Introduction}

In 2015 a few respectable experimental groups claimed that they were able to close simultaneously 
 the two basic loopholes in the Bell-test experiments \cite{B} (see also \cite{GEN} for review and references herein)  the locality loophole and the fair sampling loophole, see
 \cite{B1}-\cite{B3}.\footnote{Of course, the debate on loopholes can be continued for ever, see, e.g., the papers on signaling 
 loophole \cite{Adenier}-\cite{KU3}.} These successful experiments may lead to the conclusion that the debate between Einstein \cite{EPR} and Bohr 
 \cite{BRR} can be considered as finished and that the position of  Einstein, Podolsky, and Rosen is not supported by the experimental evidence, see, e.g., \cite{A1}-\cite{A2}. 
However, not everybody agreed  with such finalizing views \cite{KU1}, \cite{AB}. 
Personally I think that the main foundational problems related to Bell's argument have not yet been resolved and that further intensive theoretical and experimental 
studies are needed \cite{AB}. 

In this note we would like to discuss coupling of the Bell argument with the hypothesis on ergodicity of  quantum mechanics.  
The main claim is that experimental violations of the Bell inequality need not necessarily suggest to reject local-realism,  but rather ergodicity.

In physics {\it nonergodicity} (in particular, violation of the law of large numbers) is  considered as a kind of pathology.  Appearance of statistical 
data which deviate from ergodic behavior is considered as a sign that this experiment was not been well 
performed. However, this heuristics came from probabilistic analysis of macroscopic phenomena. In principle, there is no guarantee that it should work as well for 
quantum phenomena. The hypothesis about violation of ergodicity for such phenomena was actively discussed  by   Buonomano  \cite{1} since 1980th. Honestly saying,
his works did not attract so much attention. The only experiment testing the hypothesis on nonergodicity was performed in neutron 
interferometry by Summhammer \cite{2}.  This experiment can be considered as rejecting  this hypothesis and confirming ergodicity 
of quantum statistical data.\footnote{And this experiment was practically forgotten. I got to know about it
only from private conversations with J. Summhammer during his visits to V\"axj\"o - as one of curious things 
which he did when he was young. For example, by giving at least 10 talks about neutron interferometry  at the V\"axj\"o conferences H. Rauch had never mentioned 
this experiment which was performed at his institute.}   However, it  definitely cannot be considered as a decisive experiment rejecting completely the hypothesis on 
nonergodicity of quantum mechanics.  New experiments are badly needed. In  \cite{3, 4} I proposed a few experimental tests for Buonomano's hypothesis.\footnote{The theoretical 
background of these experimental proposals is $p$-adic generalization of theory of probability. We also point to the recent work of Palmer \cite{TNP}
using $p$-adic number-theoretic approach to analysis of the Bell argument.} They were even 
discussed at Atominstitute in Vienna, but they have never been performed. Typically other experiments were considered as having higher priority.
 
In preprint \cite{ERG} there was pointed out that  Buonomano's hypothesis  is related to Bell's argument. In fact, if  Buonomano were right, then it would be impossible 
to derive the Bell inequality. From this viewpoint, observed experimentally violations of this inequality need not be interpreted as the crucial argument  to reject local realism.
These violations can be treated as a sign that quantum mechanics is nonergodic. The aim of the present paper is to attract attention of the quantum community to the nonergodic 
hypothesis and to the possibility to interpret the recent exciting loophole free Bell-experiments as simply supporting  this hypothesis. We remark that in fact this hypothesis (but without 
explicit operating with the notion of (non)ergodicity) was discussed in connection to the Bell inequality in few papers, e.g., \cite{BF}, \cite{HDR}.  
 
Besides the philosophic and mathematical messages, this paper delivers the important message for experimenters: {\it to perform a series of new experiments to test 
nonegodicity in quantum physics - experiments for the straightforward tests of nonegodicity, a la  Summhammer.}

In sections \ref{ERG1}, \ref{ERG2} (which are more mathematical than the rest of the paper)  
we discuss coupling of the Bell argument with theories of ergodic stochastic processes \cite{SH} and dynamical systems \cite{AN}. 
This theories imply that dynamics behind joint measurements on components 
of a compound quantum system $S$ need not be ergodic, in spite of ergodicity of the dynamics 
behind measurements on  the subsystems $S_j, j=1,2,$  of $S.$ We connect this problem with the property of {\it weak mixing} for ergodic dynamical systems \cite{AN}.   

\section{Ergodicity: coincidence of time and ensemble averages}

We recall that if  a stationary stochastic process $x(t, \lambda)$ is   {\it ergodic} (see section \ref{ERG1}),  then the following equality holds for a sufficiently rich class of  functions $f$ 
\begin{equation}
\label{ER} E f(x) \equiv \int_\Lambda f(x(t, \lambda)) d\rho(\lambda)=
E_{\rm{time}}f(x) \equiv \lim_{n \to
\infty}\frac{f(x(t_1,\lambda)) +...+ f(x(t_n,\lambda))}{n}.
\end{equation}
Here $\rho$ is a probability measure (a process is ergodic with respect to some probability measure). The symbol $\Lambda$ denotes the space of chance parameters.
In physical modeling related to the Bell inequality, these are hidden parameters; in classical probabilistic axiomatics, see appendix, these are so called elementary events.
In particular, {\it the law of large numbers} holds true: 
\begin{equation}
\label{ERi} E x \equiv \int_\Lambda x(t, \lambda) d\rho(\lambda)=
E_{\rm{time}}x \equiv \lim_{n \to
\infty}\frac{x(t_1,\lambda) +...+ x(t_n,\lambda)}{n}.
\end{equation}
In fact, ergodicity theory  was developed to generalize of the law of large numbers.

 Thus, for an ergodic process,   the ensemble average $Ex = \int_\Lambda x(t, \lambda)
d\rho(\lambda)$ (it does not depend on $t,$ since the process is
stationary) can be approximated with an arbitrary precision by the time-average  with respect to almost any realization $\lambda,$ 
see sections \ref{ERG1}, \ref{ERG2}  for the mathematical formulations.

We emphasize  that {\it in all quantum experiments there are calculated the time averages.} To calculate the ensemble average,  one has; for example,  to be able to prepare for the fixed instance of 
time an ensemble of quantum systems and to perform the simultaneous measurement on all these systems in a way that one measurement would not disturb another. Of course, such 
an experiment cannot be performed in reality. Therefore one can use the same experimental equipment and a source which generates systems one by one, but at least experimenters
have to destroy possible effects of the previous trails, cf. Summhammer \cite{2}, see also \cite{3, 4}. However, to perform such ``loophole free'' experiments is a complicated problem.
Each experimental arrangement contains many components and in principle each of them can generate temporal correlations in data leading to violation of ergodicity. (In particular, 
Summhammer \cite{2} destroyed possible memory effects only in one component of the neutron interference experiment. However, sequential dependence leading to difference between
the time and ensemble averages can be generated by other components of the experimental arrangement, even by the source of neutrons.)  As was pointed out, it is impossible to consider 
an ensemble of experimental arrangements and to prepare a single system in each of them and then to perform the measurement on it. It seems to be reasonable to use temporal relaxation 
to destroy memory effects in all components of the experimental arrangement: to prepare just one quantum system at each interval of time $\Delta t$ and to perform the measurement on it.  
 This interval should be sufficiently  large to guarantee disappearance of the memory effects.

\section{Quantum (non)ergodicity?}

In quantum mechanics we consider the chance parameter $\lambda$
labeling runs of experiments for measuring some quantum observable
$x$ for systems prepared in the state $D.$ Thus for any run $\lambda$
we obtain a discrete process $x(t_1,\lambda),..., x(t_n,\lambda),...$
(results of measurements of $x).$ We remark that $x\equiv
x^D(t),$ so it depends on the state  $D.$ In quantum formalism it is assumed
that $x$ and $D$ are represented by self-adjoint operators,
moreover, $D$ is positively defined and it has the unit trace (a density operator representing a quantum state). The theoretical
quantum average is given by the von Neumann trace formula:
$\langle x\rangle_D=\rm{Tr} \; D x,$
which is the straightforward consequence of the Born rule determining quantum probabilities.

In a huge number of experiments there was demonstrated (long before
Bell's proposal) that the quantum average coincides with the time-average:
\begin{equation}
\label{ER1}
 \langle x\rangle_D = E_{\rm{time}}x,
\end{equation}
so
$$
\rm{Tr} \; D x =\lim_{n \to
\infty}\frac{x(t_1,\lambda)+...+x(t_n,\lambda)}{n}.
$$
However, besides Summhammer's experiment \cite{2}, there were no
experimental results confirming quantum ergodicity, namely,
coincidence of the time average $E_{\rm{time}}x$ (and hence the
quantum average $\langle x\rangle_D)$ and the ensemble average $E x.$

\section{Nonlocality and nonergodicty?}

There were no doubts that the equality (\ref{ER1}) should hold true
even for observables considered in the EPR-Bohm experiment. One of
{\it the implicit inventions of J. Bell was assuming ergodicity of quantum
mechanics and, hence, the coincidence of ensemble and time  averages,}
see (\ref{ER}). Under such implicit assumption he could  identify
theoretical quantum averages (coinciding with time averages)\footnote{The latter assumption is based 
on just the experimental experience.} with ensemble averages 
with respect to the distribution of hidden variables.

The logic of such identification can be presented as  follows: 
\begin{itemize}
\item quantum averages are well approximated by time averages;
\item  the latter coincide (through ergodicity) with ensemble averages; 
\item therefore quantum averages are well approximated by ensemble averages.
\end{itemize}
In this framework Bell derived his inequality (which is an inequality for ensemble averages) and came to the
conclusion that the quantum formalism is incompatible with local
realism. 

The crucial point of this derivation is the possibility to use 
the measure-theoretic representation of quantum correlations:
\begin{equation}
\label{LC}
\langle x_\theta \; y_\phi\rangle = 
\int_\Lambda  x_\theta(t, \lambda)  y_\phi(t, \lambda)  d\rho(\lambda).
\end{equation}
Here $\theta$ and $\phi$ are orientations of polarization beam splitters in labs 1, 2 (separated by some distance) 
and  $x_\theta,   y_\phi$ are the corresponding quantum observables, the polarization projections. (We remind that the process is assumed to be stationary.) 

The intermediate step is based on the implicit  assumption of ergodicity:
\begin{equation}
\label{LCa}
\int_\Lambda  x_\theta(t, \lambda)  y_\phi(t \lambda).  d\rho(\lambda)
= \lim_{n \to \infty} \frac{1}{n} \sum_{i=1}^n x_\theta(t_i, \lambda)  y_\phi(t_i, \lambda),
\end{equation}
However, it has never  been explicitly emphasized, neither by Bell not by thousands of his followers.   

We now remark that if the stochastic process induced by measurements of polarizations of {\it pairs of entangled photons}  
is nonergodic, then there is no reason to identify the time and ensemble averages and hence no reason to identify the ensemble and quantum averages.

 Our emphasis on ``pairs''  has an important mathematical background in theory of dynamical systems and especially those violating the condition 
of weak mixing, see section \ref{ERG2}. The latter condition is directly related to quantum correlation experiments of the Bell type. It guarantees 
ergodicity not only of the component processes, but also of their pair which is used to calculate the correlation between the components. As we know from ergodicity 
theory (section \ref{ERG2}), the condition of weak mixing can be violated for natural physical systems. In such a case the violation of ergodicity can be found only in 
experiments with compound systems. 

Nowadays the hypothesis on quantum ergodicity has been confirmed just  in the single experiment \cite{2}. In this situation the use of the ergodicity assumption 
in derivation of the Bell inequality and its generalizations  is really ad hoc.  

We remark that there were presented some proofs of the Bell-type inequality which were not based on identification of the time and ensemble 
averages \cite{ES1}-\cite{ES4}. However, these proofs are heavily based on  counterfactual reasoning (although the authors did not stress the latter 
as the crucial part of their reasoning). 

Consider two experiments on measurement of  photon polarizations for two pairs of angles $(\theta, \phi)$ and $(\theta, \phi^\prime).$ They generate two pairs of two dimensional stochastic processes,
$(x_\theta(t), y_\phi(t))$ and $(x_\theta(t), y_{\phi^\prime}(t)).$ Consider experiments with $N$ trials, for the first  setting, at the moments $t_1,..., t_N$ and, 
for the second one,     at the moments $s_1,..., s_N.$ We get the following statistical samples 
$$
(x_\theta(t_1), y_\phi(t_1), ...., (x_\theta(t_N), y_\phi(t_N))
$$ 
and 
$$
(x_\theta(s_1), y_{\phi^\prime}(s_1)),..., (x_\theta(s_N), y_{\phi^\prime}(s_N)).
$$
Then those who ``prove'' the Bell inequality without appealing to the hypothesis of ergodicity identify the statistical samples 
$$
x_\theta(t_1), ...., x_\theta(t_N)
$$ 
and 
$$
x_\theta(s_1),...., x_\theta(s_N).
$$
Here the counterfactual reasoning is involved. However, it is really impossible to combine peacefully such reasoning with the views of founders of quantum theory, N. Bohr and W. Heisenberg.
Consideration of ``experimental data'' without relation to the concrete experimental context is really against the basic principle of the quantum theory - the principle of complementarity.

Such an approach was criticized by many authors, see, e.g.,  \cite{KHR_INT}- \cite{HPL7}. In particular, in \cite{KHR_INT} there were shown that for real stochastic processes, i.e., without aforementioned identification, 
operation with solely time averages, without identification of them with the ensemble averages,  does not lead to the Bell inequality, but to its modification, i.e., the classical ensemble 
bound 2 is perturbed by the additional term. 
    
\section{Can nonergodicity save realism?}

The above discussion about the fundamental role of the implicit assumption on quantum ergodicity in derivation of the Bell inequality is 
important for better understanding of the problem of (in)compatibility of realism and quantum theory. In a nonergodic model violation of Bell's
inequality does not contradict to existence of the probability distribution of hidden variables. This was the position of De Broglie \cite{DB} in his critical 
analysis of the Bell argument against the local realism \cite{AB}.  In fact, this position is typical for those accepting the ontic-epistemic methodology 
in science \cite{ATM}. There are two levels of description: 
\begin{itemize}
 \item ontic,
\item epistemic. 
 \end{itemize}
The latter description represents our knowledge about natural phenomena; the former description is about these phenomena as they are. 
Of course, at both levels we operate with mathematical models. Epistemic models describe observations. In particular, the quantum model 
is epistemic. Ontic models describe features of natural phenomena which are not approachable through observations. It seems that 
some philosophers treat the ontic description as representing ``objective reality'' as it is. However, this viewpoint does not match with 
the mathematical modeling approach to science. Therefore it seems to be more natural to speak not about the ontic  and epistemic levels of description,
but theoretical and epistemic (observational) models. This viewpoint was advertised  by Hertz, Boltzmann, and later by Schr\"odinger, see, e.g., \cite{SH}, \cite{AB}.
Theoretical model need not be rigidly coupled to the observational. In this note we speculate that the quantum
model describes in fact nonergodic behavior and the basic structure of the theoretical (``ontic'') description, the probability distribution of hidden variables, 
need not be recovered from the time averages given by the quantum formalism. 

\section{Mathematical considerations: from separate ergodicity to compound  nonergodicity}
\label{ERG1}

Now we recall the notion of ergodicity for stochastic processes \cite{SH}.  We start with the definition of a {\it wide-sense stationary stochastic process.} Such a  stochastic process, 
denoted by $x(t),$ has constant expectation value:
$\mu _{x}=E[x(t)],$ and its  autocovariance function 
$C_{x}(\tau )=E[(x(t)-\mu _{x})(x(t+\tau )-\mu _{x})],$
depends only on the shift  $\tau$  and not on the time variable $t.$  
We stress that the expectation and autocovariance function are understood as ensemble averages, not time averages, i.e., 
$\mu _{x} = \int x(t, \lambda) d \rho(\lambda)$ and $C_{x}(\tau )=\int [(x(t, \lambda)-\mu _{x})(x(t+\tau,\lambda) -\mu _{x})]  d\rho(\lambda).$

The stationary stochastic process $x(t)$  is said to be {\it mean-ergodic} (or mean-square ergodic) if the time average 
$$\langle x \rangle_{\rm{time}}(T; \lambda) =\frac {1}{T} \int _{0}^{T} x(t, \lambda)\,dt$$ converges in squared mean to the ensemble average  $ \mu _{X}$  as $T\rightarrow \infty.$ 
The latter means that 
\begin{equation}
\label{SQR}
\lim_{T\to \infty} \int \Big\vert \frac {1}{T} \int _{0}^{T} x(t, \lambda)\,dt - \mu_x\Big\vert^2 d\rho(\lambda) =0.
\end{equation}
The $L_2$-convergence is not precisely the type of convergence which we need to work with the experimental statistical data,. Here we need convergence 
almost everywhere, i.e., for almost all $\lambda, $ 
\begin{equation}
\label{SQR0}
\lim_{T\to \infty}  \frac {1}{T} \int _{0}^{T} x(t, \lambda)\,dt = \mu_x .
\end{equation}
This is a more advanced area of the mathematical studies about ergodicity. 

Further we consider discrete stochastic processes (with sums, instead of integrals). Here  (\ref{SQR0}) has the form:
\begin{equation}
\label{SQR1}
\lim_{n\to \infty} \frac {1}{n } \sum_{i=1}^n  x(t_i, \lambda) =  \mu_x,
\end{equation}
for almost all $\lambda.$ 
The most known and widely used result of such type was obtained by Kolmogorov for independent equally distributed random variables $x(t_i, \lambda).$ This is 
{\it the Kolmogorov strong law of large numbers.} Therefore the simplest test of ergodicity of statistical data obtained in quantum experiments 
is checking the independence of trials. If they are independent, the time and ensemble averages coincide and, hence, the quantum mechanical averages coincide 
with the ensemble averages,   see \cite{KDR} detailed analysis of this problem.   However, the assumption of independent trials is too strong.

 The general ergodic theorem 
is known as {\it Birkhoff-Khinchin theorem.} For its formulation, we need a stronger notion of stationarity based on consideration of joint probability distributions and the 
corresponding notion of ergodicity.
A stochastic process $x(t), t \in \mathbf{N}=\{1,...., n,...\},$ is called stationary if 
$$
P( (x(1), ..., x(n)) \in A) = P( (x(1+m), ..., x(n+m)) \in A), 
$$ 
for any $m$ and $n$ and any Borel subset $A$ of $\mathbf{R}^n.$ If a process is stationary, then it is, of course, stationary in the wide-sense, but not vice versa. 

Consider the space of all infinite sequences of real nu
mbers $$\Omega = \mathbf{R}^\mathbf{N}=\{x=(x_1,...,x_k,...)\}$$ endowed with the $\sigma$-algebra ${\cal F}$ generated by cylindric 
subsets, i.e., subsets of the form $A_{i_1 i_2 ... i_n}= \{ x\in   \mathbf{R}^\mathbf{N}: (x_{i_1} x_{i_1} ... x_{i_n}) \in A\},$ where  $A$ is a Borel subset of $\mathbf{R}^n.$   Consider 
the probability measure $P$ on this $\sigma$-algebra determined by finite dimensional probability distributions of the process $x(t).$ 

In the space $\Omega$ define the left-shift operator
$$T (x_1....x_n....)= T (x_2....x_n....).$$ Then the process is stationary if and only if $P(T^{-1}({\cal A})) = P({\cal A})$ for any${\cal A} \in {\cal F}.$

An event     ${\cal A} \in {\cal F}$ is shift-invariant if $T^{-1}({\cal A})= {\cal A}.$  The process $x(t)$ is called {\it ergodic}
 if every shift-invariant event ${\cal A}$ is trivial, i.e., $P({\cal A})=0$ or 1. 

{\bf Theorem 1.} (Birkhoff-Khinchin) {\it Let the stochastic process $x(t), t=1,2,...,$ be stationary and ergodic. If  $E \vert x(1) \vert< \infty,$ then 
(\ref{SQR1}) holds almost everywhere.}

Now we consider an interesting problem of theory of ergodicity which has close relation to the identification of the ensemble averages with the time (and hence quantum-theoretical)
 averages in the  Bell framework. Consider two  stationary ergodic processes $x(t)$ and $y(t).$ Is the process $z(t)= (x(t), y(t))$ ergodic? The answer is `no'.  We remark that this 
implies that the product $x(t)y(t)$ of two stationary ergodic processes need not be ergodic.  Thus in principle
in quantum theory nonergodicity can be generated by measurements on compound systems. At the same time measurements 
of each system separately generate ergodic stationary processes. In such a case measurements on  compound systems  really 
have the special feature, nonergodicity. We can speculate that nonergodicity may be really a  delicate issue. It can happen that, for each of  stochastic processes $x_\theta(t)$ and $y_\phi(t)$  generated 
by measurements of the polarization observables  in labs 1 and 2, the ensemble, time, and  quantum mechanical averages coincide. But at the same time, for correlations,  time and ensemble 
averages can be different, since  the product $x_\theta(t) y_\phi(t)$  of these processes 
need not be ergodic. However, to derive the Bell inequality one has to operate with correlations and hence (ensemble) averages of products of stochastic processes.

\section{Mathematical considerations: coupling to theory of dynamical systems, weak mixing}
\label{ERG2}

Turn to the representation of a stochastic process $x(t)$ by the probability distribution $P$ on the space of trajectories 
of this process,  $\Omega = \mathbf{R}^\mathbf{N}$ endowed with the $\sigma$-algebra ${\cal F}$ generated by cylindric 
subsets. The iterations of the  left shift map $T$ can be treated as a discrete dynamical system. (We recall that we consider 
only discrete time processes, $t=1,2,...\;.)$ Thus, instead of the processes, we can work with the dynamical system 
$(\Omega, {\cal F}, P, T).$  Stationarity of the  stochastic process is equivalent to measure preserving property of $T.$ 

In turn, we can start with an arbitrary dynamical system  \cite{AN} $(X, {\cal G}, \mu, T),$ where $\mu$ is an arbitrary probability measure on the $\sigma$-algebra ${\cal G}$
and $T: X \to X$ is a measurable map.
We shall consider measure preserving dynamical systems. It is
 {\it ergodic} if every $T$-invariant event ${\cal A}\in {\cal G}$ is trivial, i.e., $\mu({\cal A})=0$ or 1.

Consider now a measurable function $f: X \to \mathbf R.$ Such functions represent observables.  Set $x_f(t; \lambda) = f (T^t \lambda), \lambda \in X.$ 
Then $x_f(t)$ is a stochastic process. 

Thus we can speak either about stochastic processes or dynamical systems. For latter, we formulate the famous Birkhoff theorem:

{\bf Theorem 2.}  {\it Let $(X, {\cal G}, \mu, T)$ be an ergodic dynamical system. Then, for function 
$f,$  such that  $E \vert f \vert <\infty,$ 
\begin{equation}
\label{SQR7}
\lim_{n \to \infty} \frac {1}{n} \sum_{i=1}^n f(T^i \lambda)   =  \int_X f(\lambda) d \rho(\lambda) .
\end{equation}
almost everywhere.}

In particular, if we consider the dynamical system $(\Omega, {\cal F}, P, T)$ corresponding to the stochastic process $x(t),$ then we can select $f(x_1,...., x_n, ...)=x_1.$ 
For this map, we have $x_f(t; \lambda)= x(t; \lambda).$ 

However, for many purposes, the ergodicity constraint is too weak and typically there are considered dynamical systems satisfying stronger constraint known 
as {\it mixing.} A dynamical system  $(X, {\cal G}, \mu, T)$ is called  {\it strongly mixing} if, for any $A , B \in {\cal G},$ 
 $$
\lim_{n \to \infty} \mu(A \cap  T^{- n} B ) = \mu ( A ) \mu (B),
$$
where $T^{- n} B \in {\cal G},$ $T$ is measurable.
\footnote{ Now by ignoring the problem of measurability consider, instead  of $T$-preimages,  direct $T$-images. Then by taking into account that $\mu(X)=1,$ we have: 
$$
\lim_{n \to \infty} \frac{ \mu (T^{n} A \cap   B )}{ \mu ( B )} = \lim_{n \to \infty} \mu (T^{n} A \vert   B ) = \mu( A )= \frac{ \mu( A )}{ \mu( X)}.
$$ 
In a shaker there are two liquids, wine and water, initially wine was concentrated in the volume $A.$ So, what is the proportion of wine in an arbitrary 
volume $B$ of the shaker? It is the same as the original proportion of wine in the shaker, so it is doing really good mixing. }
For our aim, it is enough to consider a weaker property.  A dynamical system is said to be {\it weak mixing} if 
$$
   \lim _{n\to \infty }{\frac {1}{n}}\sum _{k=0}^{n-1}\left|\mu (A\cap T^{-k}B)-\mu (A)\mu (B)\right|=0.
$$
Strong mixing implies weak mixing which in turn implies ergodicity. However, inverse is not valid.  

{\bf Theorem 3.} {\it  A dynamical system $(X, {\cal G}, \mu, T)$  is weakly mixing  if and only if,  for any ergodic dynamical system $(Y, {\cal G}, \nu, Q),$ 
the direct product of these dynamical systems is also  ergodic.}

Thus in our analysis of the Bell argument a violation of the Bell inequality can (but need not) be coupled to 
violation of weak mixing condition.

The evident objection to this kind of reasoning is that, for the Bell states, it is natural to assume that  stochastic processes observed in lab 1 and lab 2 are based on the 
same dynamical system $(X, {\cal G}, \mu, T)$ and one can play only with selection of functions $f_1$ and $f_2$ determining observables. Of course, matching of  the latter assumption 
to the real experimental situation is not eveident at all. But we shall not go in such a discussion about matching. In fact, mathematics gives us the formal argument  implying 
that even identification of two dynamics (determining the temporal structure of stochastic processes observed in lab 1 and lab 2) 
does not change the previous consideration about coupling of weak mixing with a violation of the Bell inequality.    

{\bf Theorem 4.} {\it  A dynamical system $(X, {\cal G}, \mu, T)$  is weakly mixing  if and only if its
the direct product with itself is also  ergodic.}
 
 Thus even if the dynamics of  each of entangled photons in the Bell experiment is described by the same dynamical map $T^t$ which is ergodic by itself, then 
 the dynamics  of the pair of  these photons need not be ergodic (if this dynamical  map is not weak mixing). I agree that there is a kind of mystery in disappearance of ergodicity 
 for a compound system. However, this is not physical, but mathematical mystery. It was discussed a lot in mathematical literature devoted to theory of stochastic 
processes and dynamical systems. There is no possibility to go deeper into the corresponding  mathematical discussion in this paper. So, further analysis of consequences of coupling 
of the Bell argument and spurious quantum nonlocality with theory of mixing and ergodicity will be presented in a future paper.  In particular, if the hypothesis on violation 
of ergodicity for processes generated by measurements on compound quantum systems is correct, then we have to understand coupling between entanglement and violation of 
weak mixing. This is the exciting problem.

Finally, we remark that coupling of the theory of dynamical systems with interference phenomena and 
violation of Bell's inequality was considered in papers of a few authors, see, e.g., \cite{DB}, \cite{Palmer7}. 

\section{Concluding remarks} 

A violation of the Bell-type  inequalities predicted by the quantum formalism and confirmed by a series of experiments need not be interpreted as incompatibility 
of local realism and quantum theory; they can be as well interpreted as evidences of  a violation of ergodicity for quantum measurement process. Under the assumption 
of nonergodicity one can proceed without rejecting realism. However, the latter is closely coupled to structuring physics through two levels of description of physical 
phenomena, ontic-epistemic,  or two mathematical models, theoretical-observational.  The hypothesis on quantum nonlocality (spooky action at the distance) can be rejected 
either as an absurd assumption (see Einstein, Podolsky, and Rosen \cite{EPR}) or by using Occam's Razor.

 We remark that GHZ-like tests, which in some aspects present a stronger claim for nonlocality than Bell's arguments, can also be analyzed from the viewpoint of violation of the hypothesis 
of ergodicity. Such analysis is based on the contextual representation of these tests \cite{GHZ}. However, this is a separate and complicated problem which deserves special consideration.

We finally make the following remark about {\it possible future experiments in quantum foundations.}  It would be natural to
switch the experimental research in quantum foundations from the study of  probabilistic behavior of  outputs of measurements for entangled system to experiments that
can test  ``pure'' (not mixed with nonlocality) ergodicity. The next step is testing the hypothesis on weak mixing, but not through consideration of compound systems 
and the the pairs of processes generated by measurement on the sub-systems. In fact, weak mixing can be checked straightforwardly for noncompound systems. (However, for the 
moment I do not have the real experimental proposals.) 

\bigskip

This paper was partially supported by the EU-project Quantum Information Access and Retrieval Theory (QUARTZ), Grant No. 721321.

\end{document}